\documentclass[aps,prd,twocolumn,showpacs,nofootinbib,amsmath,amssymb]{revtex4}
\usepackage{graphicx}
\usepackage{dcolumn}
\usepackage{bm}
\usepackage{hyperref}
\begin{document}

\title{Stochastic resonance with matched filtering}

\author{Li-Fang Li}
  \affiliation{Department of Physics, Beijing Normal University, Beijing 100875, China}

\author{Jian-Yang Zhu}
\thanks{Author to whom correspondence should be addressed}
  \email{zhujy@bnu.edu.cn}
  \affiliation{Department of Physics, Beijing Normal University, Beijing 100875, China}

\date{\today}
\begin{abstract}
Along with the development of interferometric gravitational wave
detector, we enter into an epoch of gravitational wave astronomy,
which will open a brand new window for astrophysics to observe our
universe. Almost all of the data analysis methods in gravitational
wave detection are based on matched filtering. Gravitational wave
detection is a typical example of weak signal detection, and this
weak signal is buried in strong instrument noise. So it seems
attractable if we can take advantage of stochastic resonance. But
unfortunately, almost all of the stochastic resonance theory is
based on Fourier transformation and has no relation to matched
filtering. In this paper we try to relate stochastic resonance to
matched filtering. Our results show that stochastic resonance can
indeed be combined with matched filtering for both periodic and
non-periodic input signal. This encouraging result will be the first
step to apply stochastic resonance to matched filtering in
gravitational wave detection. In addition, based on matched
filtering, we firstly proposed a novel measurement method for
stochastic resonance which is valid for both periodic and
non-periodic driven signal.
\end{abstract}


\maketitle

\section{introduction}

Since the concept of stochastic resonance (SR) was originally put
forward in the seminal papers by Benzi and others \cite{benzi}, it
has continuously attracted considerable attention \cite{review}. SR
was subsequently found to occur in many kinds of nonlinear systems
\cite{bezrukov}, even linear systems with multiplicative noise
\cite{berdichevsky} and diverse fields. Not only classical system
but also quantum system \cite{quantum} show SR behavior. Generally,
SR emerged as a paradigm whose universal character was shown to be
intimately related to the pervasiveness \cite{dykman} of the
fluctuation-dissipation theorem \cite{landau}. The possible use of
SR in connection with weak signal detection experiments
\cite{jung91,gong92,inchiosa,loerincz96,galdi98a,galdi98b}, with
special reference to gravitational wave detection, was suggested
since the infancy of SR \cite{gammaitoni89}. The application of
stochastic resonance to gravitational wave detection includes two
aspects, one is data analysis \cite{galdi98a}, and the other is
related to detector itself \cite{karapetyan04}. As to data analysis
method, the research about stochastic resonance almost all use
Fourier transformation method to deal with the data with noise. On
the other hand, the data analysis method in gravitational wave
detection community is mainly the so called matched filtering
method, i.e. maximizing the likelihood ratio over the parameters of
expected signal templates \cite{thorne87,cutler94,owen99,babak}. So
it is interesting to investigate whether we can combine the advanced
data analysis method used in gravitational wave detection, matched
filtering, with stochastic resonance. In this work, we will study
whether the stochastic resonance phenomena can be observed also with
matched filtering method to do data analysis. Our work shows that
stochastic resonance does emerge with matched filtering data
analysis method for periodic signal, corresponding to the periodic
gravitational waves; chirp signal, corresponding to the
gravitational wave radiated from merging black holes or neutron
stars and even random pulse signals.

Till now we have three kinds of methods to characterize stochastic
resonance. The first one is power spectrum include output signal
power itself and the ratio of output signal power to noise
background. This kind of measurement method is most convenient and
popular when dealing with stochastic resonance. But the disadvantage
of this method is that it can only deal with periodic stochastic
resonance. The second method is residence time distribution
\cite{lofstedt94}. The third one is information theoretic measures
which includes the Shannon transinformation rate \cite{heneghan96};
the Fisher information \cite{greenwood00}; Kullback, Shannon and
other kinds of entropies \cite{entropy}; and information-theoretic
distance measures \cite{robinson98,collins95}. In this aspect, our
work shows that matched filtering method can be looked as a novel
measurement method to characterize stochastic resonance. The method
of matched filtering is very robust that it can characterize both
periodic and non-periodic stochastic resonance.

This paper is organized as following. In section II, we will give
a brief review of matched filtering method used in gravitational
wave data analysis. Then we present out our model equations used
in this paper in section III. In section IV, we give out our
result on stochastic resonance with matched filtering method. At
last we discuss our result and its implications in section V.
\section{brief review of matched filtering techniques used in gravitational wave detection}

First we flesh out the brief description of matched filtering
techniques used in gravitational wave detection \cite{thorne87}.
Assume our data stream takes form like
\begin{eqnarray}
h(t)=s(t)+n(t),
\end{eqnarray}
where $s(t)$ is signal, while $n(t)$ is Gaussian white noise. The
likelihood ratio is defined as \cite{cutler94,owen99}
\begin{eqnarray}
{\cal M}\equiv\max_{t_c}\frac{\langle
h,u(t-t_c)\rangle}{\text{rms}\langle u,u\rangle}.
\end{eqnarray}
Here $u$ is the template determined by parameter in some space used
to filter the data stream $h(t)$. The maximal value is taken with
respect to the time delay $t_c$ of the template. When the signal is
periodic or almost periodic the maximal value can be equivalently
taken respect to the initial phase of template. And the inner
product is given by
\begin{eqnarray}
\langle a,b\rangle&\equiv& 4
\text{Re}\left[\int_0^\infty\frac{\tilde{a}^*(f)\tilde{b}(f)}{S_n(f)}df\right]\nonumber\\
&=&2\int^\infty_0\frac{\tilde{a}^*(f)\tilde{b}(f)+\tilde{a}(f)\tilde{b}^*(f)}{S_n(f)}df.
\label{inner_product}
\end{eqnarray}
It is the noise-weighted cross-correlation between $a$ and $b$. Here
$\tilde{a}(f)$ stands for the Fourier transform of $a(t)$
\begin{eqnarray}
\tilde{a}(f)\equiv\int_{-\infty}^\infty e^{i2\pi ft}a(t)dt,
\end{eqnarray}
and $*$ stands for complex conjugate. $S_n(f)$ is the one-sided
power spectral noise density of the noise stream $n(t)$
\begin{eqnarray}
S_n(f)\equiv 2\text{E}[|\tilde{n}(f)|^2],
\end{eqnarray}
where E[  ] denotes the expectation value over an ensemble of
realizations of the noise. Since our data are all real, we can use
one-sided power spectral instead of two-sided one. Above description
is in frequency domain. In time domain, the inner product
(\ref{inner_product}) takes form \cite{cutler94}
\begin{eqnarray}
\langle a,b\rangle\equiv
\int_{-\infty}^\infty\int_{-\infty}^\infty
a(t)w(t-\tau)b(\tau)dtd\tau,
\end{eqnarray}
where $w(t)$ is Wiener's optimal filter \cite{kantz97}, which can
be approximated as
\begin{eqnarray}
w(t)\approx\int_{-\infty}^\infty e^{-i2\pi ft}\frac{1}{S_n(f)}df.
\label{wiener}
\end{eqnarray}
Then the task of data analysis in gravitational wave detection is
to maximize the likelihood ratio ${\cal M}$ over the parameters
space of expected signal templates and then fix a detection
threshold on this statistics. The detection threshold is simply
established by the desired level of false alarm probability. The
maximum likelihood is compared with the aforementioned threshold
and a detection is announced whenever the threshold is crossed.

In most pragmatic cases, the assumption of Gaussianness and
whiteness to the noise is not valid, especially when noise passes
through a nonlinear system, such as the system considered in this
work. So the Wiener's filter (\ref{wiener}) is not valid any more.
Therefore, we adopt the matched filtering method and the idea of
Horner efficiency in optics \cite{caulfield82} to define matched
ratio as
\begin{eqnarray}
{\cal
M}=\max_{t_c\in\Delta}\frac{\int_{-\infty}^\infty|\int_{-\infty}^\infty
h(t-\tau)u(\tau-t_c)d\tau|dt}{\int_{-\infty}^\infty|
h(t)|dt\int_{-\infty}^\infty|u(t)|dt}, \label{SNR}
\end{eqnarray}
where $h(t)$ is the data stream and $u(t)$ is the template as
above. $t_c$ is the time delay for the template. $\Delta$ is the
time delay range where the delay time is taken from. Theoretically
$\Delta=(-\infty,+\infty)$, but it is impossible to take this
infinity range practically. In practice, if the system has some
characteristic time scale (such as the period of periodic system),
this $\Delta$ can be taken as this time scale directly. Otherwise,
there is an ambiguity to set this range. In the following sections
we will come back to this problem and propose one trick to solve
it. Different to equation (1), now $n(t)$ may be any kind of
noise, white or colorful, Gaussian or non-Gaussian. So our
quantity (\ref{SNR}) will be valid in more general case than
equation (2). Here the template is the signal profile we expected,
and the different expected signal profiles up to a scale are
looked as one template because we have already taken normalization
in equation (\ref{SNR}). Since we do not know when the expected
signal will begin in the data stream, we need take the maximal
value respect to the delay time $t_c$.

\section{model equation}
The model equation used in this paper is the famous overdamped
bistable system \cite{hu91}
\begin{eqnarray}
\dot{X}=aX-bX^3+S_p(t)+\Gamma(t),\label{system}
\end{eqnarray}
with
\begin{eqnarray}
\langle\Gamma(t)\rangle=0,\langle\Gamma(t)\Gamma(t')\rangle=2D\delta(t-t').
\end{eqnarray}
$a$ and $b$ are two parameters. We set $a=b=1$ in this paper. This
system has been widely investigated in stochastic resonance
literature \cite{hu91,bistable,hu92}. But all the studies used
Fourier transformation method to do the data analysis. While in this
paper we will adopt popular method used in gravitational wave
detection---matched filtering method instead. $S_p(t)$ is the signal
which is the input signal of our system. If we take our model
equation as the toy model equation of gravitational wave detector,
$S_p(t)$ is nothing but the gravitational wave signal. If we take
our model equation as the afterward analysis tools for gravitational
wave data analysis, $S_p(t)$ is the raw data stream of the
gravitational wave detector's output, which includes signal itself
and noise. In the following we take the former viewpoint. As to
numerical method, the stochastic Runge-Kutta algorithm \cite{risken}
is used in this work to simulate equation (\ref{system}). The time
step is set as $\Delta t=0.1$.

We take $200$ systems to realize the random process and take the
ensemble average to analyze the random data stream produced by
numerical method. Then our measurement quantity of stochastic
resonance becomes the ensemble average of equation (\ref{SNR}),
i.e. $E[{\cal M}]$. More explicitly, the measurement quantity we
introduced in this work to characterize stochastic resonance is
\begin{eqnarray}
E[{\cal
M}]=\frac{1}{N}\sum_{i=1}^N\max_{t_c\in\Delta}\frac{\int_{-\infty}^\infty|\int_{-\infty}^\infty
h_i(t-\tau)u(\tau-t_c)d\tau|dt}{\int_{-\infty}^\infty|
h_i(t)|dt\int_{-\infty}^\infty|u(t)|dt}.\label{mysnr}
\end{eqnarray}
Here $h_i(t)$ is the data stream from i'th system. $N$ is the
number of the systems in the ensemble. In practice, $h$ and $u$
are two arrays of numbers. Let us assume the length of the array
is $m$. Then the convolution $\int_{-\infty}^\infty
h_i(t-\tau)u(\tau-t_c)d\tau$ can be done with standard algorithm.
And we adopted rectangular method to approximate the other
integral. We need pay more attention to the time delay problem
since our data array is finite. In practice, we take the elements
from number $\frac{t_c}{\Delta t}$ to $m$ of original discretized
data stream $u$ for delayed template, while the data stream $h$ is
correspondingly taken as the elements from number $1$ to
$m-\frac{t_c}{\Delta t}+1$ of original array. Here $\Delta t$ is
the sample time between the nearest two elements of the array
which is the time step of the numerical simulation in our work.

\section{stochastic resonance with matched filtering method}
In this section we present our main result on stochastic resonance
with matched filtering method. We find that stochastic resonance
shows up perfectly with characteristic quantity (\ref{mysnr}) no
mater what kind of input signal we use. Our input signal includes
periodic one, which corresponds to the periodic gravitational
waves; chirp signal, corresponding to the gravitational wave
radiated from merging black holes or neutron stars and random
pulse ones. All of these stochastic resonance phenomena can be
described well by the quantity (\ref{mysnr}) we first introduced
in this paper.
\subsection{periodic input}
\begin{figure}[h!]
\includegraphics[width=0.5\textwidth]{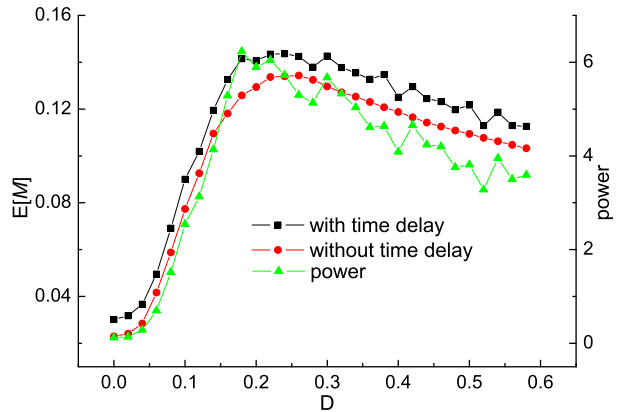}
\caption{(color online) Stochastic resonance behavior of overdamped
bistable system (\ref{system}) for periodic signal
 with matched filtering measurement method. Ensemble averaged matched ratio($E[{\cal M}]$ Eq.(\ref{mysnr}), solid line)
 and the output signal power (dashed line) are plotted. Both show perfect stochastic resonance behavior. In addition this two
 results are consistent to each other very well. The line marked with triangle is $E[{\cal M}]$ without time delay.
 We can see that $E[{\cal M}]$ with and without time
 delay show the same qualitative behavior. This is because that our template actually has no
 time delay respect to the output data stream.} \label{fig1}
\end{figure}

Firstly, we set the input signal in equation (\ref{system}) as
sinusoidal function of time
\begin{eqnarray}
S_p(t)=A\cos(\omega t), \label{periodic}
\end{eqnarray}
with $\omega=0.2, A=0.05$ closely following \cite{hu91}. This signal
can model periodic gravitational wave very well. Periodic
gravitational waves are emitted by various astrophysical sources.
They carry important information on their sources (e.g., spinning
neutron stars, accreting neutron stars) and also on fundamental
physics, since their nature can test the model of general
relativity. As to the expected profile of output signal we take
$S_p(t)$ directly as our template. In numerical simulation we evolve
$6000$ steps which equivalents to $600$ units of time which
corresponds to about $20$ periods of input signal. Since the
template here is periodic function, we can set the range $\Delta$
for the delay time $t_c$ as $(0,\frac{2\pi}{\omega})$. The numerical
result are presented in Fig.\ref{fig1}. In this figure we plot the
$E[{\cal M}]$ (solid line) together with output signal power (dashed
line) measured with traditional Fourier transformation method. The
two results are consistent to each other very well.

\subsection{chirp signal input}
Secondly, we set the input signal as the gravitational wave signal
radiated from two merging black holes or neutron stars which reads
\begin{eqnarray}
S_p(t)=\frac{A}{\sqrt[4]{t_{\text{coal}}-t}}\cos\left[\omega(t_{\text{coal}}-t)\right]^{5/8}.
\label{chirp}
\end{eqnarray}
$A$ and $\omega$ are two parameters determined by the masses of
the binary system's sub-bodies and the distance between the source
and the detector. $t_{\text{coal}}$ is the time of coalescence.
This wave form is theoretical prediction of post-Newtonian
analysis for binary systems which includes black hole-black hole
binary, black hole-neutron star binary and neutron star-neutron
star binary. In this work we set $A=0.2$, $t_{\text{coal}}=601$
and $\omega=6$. This signal is a typical wideband signal. Our
parameters setting makes the frequency band locate around $0.2$.
The same to above subsection, the simulation time is $600$ time
units which is very near the coalescence time $601$.

\begin{figure}[h!]
\includegraphics[width=0.5\textwidth]{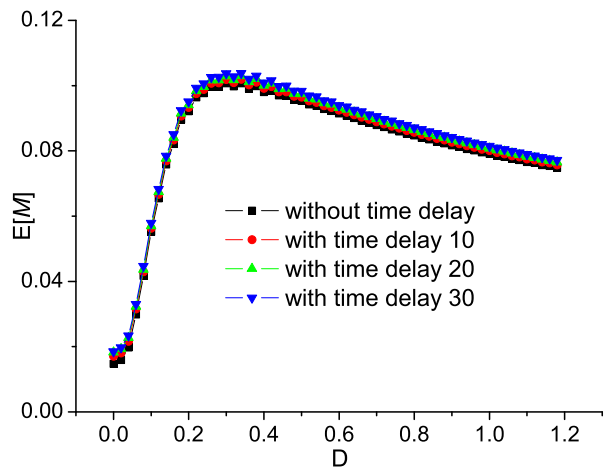}
\caption{(color online) Stochastic resonance behavior of overdamped
bistable system (\ref{system}) for chirp signal
 with matched filtering measurement method. The $E[{\cal M}]$ without time delay (marked with square) and
 with different time delay ($10$ marked with disk, $20$ marked with up triangle and $30$ marked with down triangle) are plotted respectively. All of
 these lines show well convergence behavior which implies that the one without time delay has already described the true stochastic resonance
 behavior already. This is consistent with expected result, because our input signal and template has no time delay at all with equation
 (\ref{system}).} \label{fig2}
\end{figure}

Since this signal is not periodic, we have no time scale to guide us
to choose the range of time delay $\Delta$ in equation
(\ref{mysnr}). Here we propose to use try error method to find out
the true stochastic resonance behavior. Try error method we mean
trying different time delay range from small to large till the
result converges. Here converge means resulting in the almost same
optimal noise intensity. Theoretically the time delay is longer the
convergence is better. But the data stream has finite length in
practice. These results are plotted in Fig.\ref{fig2}. Since our
input signal and template has no time delay at all, we expect the
line without time delay should be the converged result. But we see a
small but nonvanishing difference between different time delays.
This is because our data stream is some short, only 600 steps, while
different time delay results in different effective length of data
array.

\subsection{random pulse signal}
At last but not at least, we use random pulse signal as the input.
We follow \cite{hu92} closely to construct the random pulse. Each
pulse takes a value
\begin{eqnarray}
S_p(t)=A\text{ or }S_p(t)=-A
\end{eqnarray}
and persists for time $\frac{T}{2}$. Each $A$ or $-A$ are randomly
taken with probability $1/2$ respectively. In this work we set
$A=0.05$ and $T=\frac{2\pi}{0.2}$. We set simulation time $600$
units still.

\begin{figure}[h!]
\includegraphics[width=0.5\textwidth]{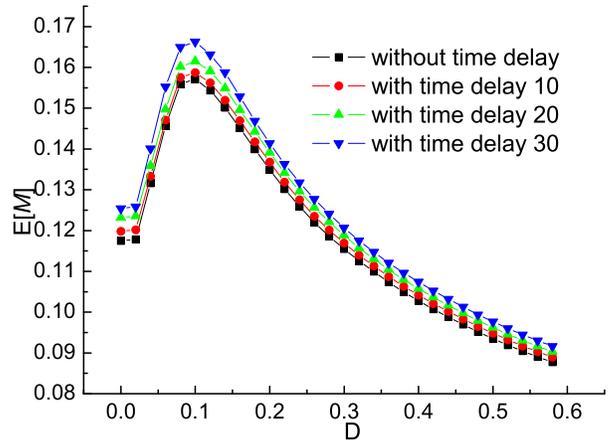}
\caption{(color online) Stochastic resonance behavior of overdamped
bistable system (\ref{system}) for
 random pulse input with matched filtering measurement method. This result is very similar with above ones but shows
 different optimal noise intensity. This is reasonable because of the different driven signal. This figure shows that matched filtering method can
 also describe non-periodic stochastic resonance very well.} \label{fig3}
\end{figure}

Similar to above chirp case, we use try error method to determine
the stochastic resonance again. We concluded all these results in
Fig.\ref{fig3}.

All three figures show that the one without time delay has already
described the true stochastic resonance behavior already. This is
consistent with expected result, because our input signal and
template has no time delay at all with equation (\ref{system}).
But generally, the time delay is important for the stochastic
resonance measurement quantity (\ref{mysnr}) when the template
does has time delay compared with the data stream.
\section{conclusion and discussion}
Since 2003 the ground-based interferometric gravitational wave
detector Laser Interferometer Gravitational Wave Observatory (LIGO)
has taken several real experimental data. And other high sensitive
detector such as VIRGO, GEO and TAMA have also achieved their
expected sensitivities. At the same time, the space-based
interferometric gravitational wave detector LISA is also under way.
Along with the development of these high sensitive detector, we are
entering into an epoch of gravitational wave astronomy.
Gravitational wave detection is a typical example of weak signal
detection where stochastic resonance may be applicable. But almost
all of the data analysis method in gravitational wave detection are
based on matched filtering. In contrast, almost all of the
stochastic resonance theory is related to Fourier transformation
which does not relate to matched filtering. Here our work do the
first step trying to relate matched filtering with stochastic
resonance.

The encouraging result of our work is that we do find that the
stochastic phenomena emerges with matched filtering data analysis
method. And more interestingly, matched filtering method can deal
with not only periodic but also non-periodic stochastic resonance.
Specifically, we show that matched filtering can deal with
stochastic resonance which is driven by sinusoidal signal, which
corresponds to periodic gravitational wave; chirp signal, which
corresponds to gravitational wave from merging binary system; and
random pulse signal. The interesting problem in the following is
whether and how to apply our findings and stochastic resonance
theory to do gravitational wave data analysis and gravitational wave
detector construction.

On the other hand the measurement quantity (\ref{mysnr}), first
proposed in this work, is a novel characteristic measurement for
stochastic resonance. Our quantity can not only deal with periodic
but also non-periodic stochastic resonance, which improves the
conventional Fourier transform method. We hope our new stochastic
resonance measurement can find its usage in the future research.

\acknowledgments The work was supported by the National Natural
Science of China (No.10875012) and the Scientific Research
Foundation of Beijing Normal University. Numerical computations were
performed on the cluster of HSCC of Beijing Normal University.

\end{document}